\documentclass[10pt,letterpaper]{article} 
\usepackage[english]{babel}

\newcommand{\beq}{\begin{equation}} 
\newcommand{\eeq}{\end{equation}}
\newcommand{\lbl}{\label}
\newcommand{\q}{\quad}

\newcommand{\re}[1]{(\ref{#1})}

\newcommand{\bE}{{\bf E}}
\newcommand{\Ebar}{\bar {E}}
\newcommand{\tauprime}{\tau^\prime}
\begin{document} 
\begin{center}
\Large
\bf
Variant forms of Eliezer's theorem\\
\normalsize
\end{center}
\section{Introduction} 
Consider a Coulomb electric field $\bE$ 
cut off at large distances $r \geq r_0$ from the source: 
\begin{eqnarray}
\lbl{largecutoff} 
|\bE (r)| &:=& 
\frac{2}{3} Q^2 \frac{1}{r^2}  \q \mbox{for $0 < r \leq r_0$} \\
&:=&  0 \q\q  \mbox{for \q $ r > r_0$} \q. \nonumber 
\end{eqnarray} 
Here $\bE$ is a radially outward  electric field 
with source at the origin.  
Also, $Q^2$ is a positive parameter, 
the square making its positivity obvious
at a glance, and the normalization factor $2/3$ is for later convenience.  

This is like the field of a proton, except that it is cut off,
so for ease of language, we'll call the source a ``proton''. 
Consider an electron traveling on the
$x$-axis which enters the field from the left at $x = -r_0$
with initial velocity $v_0 > 0$, so it's traveling radially toward the proton
from left to right.  

Physically, we expect the electron to be attracted to
a collision with the proton in a finite time.
Eliezer showed that for a full Coulomb field (i.e., not cut off),
an electron moving radially
in accordance with the Lorentz-Dirac (LD) equation  
will turn around before it reaches the proton
and thereafter travel away from the source toward $x = -\infty$ 
with acceleration increasing exponentially with proper time.  

This is true for all initial accelerations of the electron.
(Recall that to specify a unique solution to the LD equation,
one must specify an initial acceleration, along with initial position
and velocity.)
That is, {\em all} such solutions are ``runaway''.

Eliezer's original proof \cite{eliezer},
applied not only to a static Coulomb source,
but to the more nearly realistic situation of two oppositely
charged particles of identical mass (a positron and an electron, say)
moving symmetrically on a line.
For later and sometimes simpler proofs, see \cite{hsing}, \cite{hsing2}, 
\cite{baylis}, \cite{parrottbook}.

The situation is similar, but not quite the same,
for the cutoff Coulomb field.
One difference is that zero acceleration seems 
the only physically reasonable initial acceleration
when the electron enters the field.
If the electron has been moving at constant velocity
before it enters the field (as physically expected),
its acceleration at $x= -r_0$ is obviously zero.
Any other initial acceleration implies preacceleration.  
It is a basic assumption of the discussion to follow that
preacceleration is considered physically unreasonable,
so that we may assume that the electron has zero initial acceleration 
when it enters the field.  

The proofs to be given are similar to those of \cite{parrottfp},
where they were used to conclude that under the above assumptions,
all solutions are runaway. 
Here they will be applied 
to estimate how close 
the electron gets to the proton before turning around.
They were constructed to convince proponents of the LD equation
that the unphysical behavior predicted by Eliezer's theorem
(e.g., solutions ``runaway'' in the ``wrong'' direction)
can occur even in the ``classical'' regime in which the 
electron never encounters fields large enough to 
require a quantum-mechanical analysis. 

We want to find simple conditions which guarantee
that the electron will turn around before it gets within 
a prescribed minimum distance $r_1$. 
We'll show that this will occur for all sufficiently small
initial velocities.  This result does not even require that
the field be a cutoff Coulomb field---it is true for 
virtually any nonzero, spherically
symmetric field which is directed radially outward and which   
vanishes at sufficiently large distances from the origin.
\newtheorem{theorem}{Theorem}
\begin{theorem}
\lbl{thm1} 
Let $\bE = \bE(r)$ denote a spherically symmetric 
electric field at distance $r > 0$ from the origin. 
Assume that $\bE$ is directed outward from the origin 
(where it does
not vanish), 
and that $\bE$ vanishes at large distances from the origin.  
Let $r_0 $ denote the smallest distance such that $\bE(r) = {\bf 0}$ 
for $r > r_0$.  That is,  $r_0 := \inf \{\ r \ | \ \bE(r^\prime)  = {\bf 0}
\mbox{ \ for all $r^\prime > r $}\ \}$. 
Assume that for any $\epsilon > 0$, 
$$
  \int_{r_0 - \epsilon}^{r_0}  |\bE (r)|  \, dr \ > 0 
\q.
$$ 

Consider an electron moving radially in this field 
(on the $x$-axis, say)
according to the Lorentz-Dirac equation.  
Suppose that it is moving rightward with positive 
initial velocity $v_0$ when it enters the field
at $x = -r_0$,
and assume that it  
has zero proper acceleration at this time. 

Let a distance $r_1 > 0$ be given.  
If the electron's  initial velocity  $v_0 $ 
is sufficiently small,
the electron will turn around before it reaches $x = -r_1$. 

After turning,
it travels back toward $x = -\infty$ with proper acceleration
increasing exponentially with proper time. 
(The exponential increase continues even after the particle leaves the field.) 
\end{theorem} 
\noindent 
\begin{description}
\item[Remark:]
The condition 
$  \int_{r_0 - \epsilon}^{r_0}  |\bE (r)|  \, dr \ > 0 $
should be satisfied by any physically reasonable field. 
For example, it is satisfied by any continuous field which does
not vanish identically.  The condition was stated as above so that
the theorem would apply both to continuous fields and to cutoff 
Coulomb fields like \re{largecutoff}. 
\end{description}

Later we'll modifiy the proof of this theorem to show
that for a cutoff Coulomb field \re{largecutoff},
given any $r_1$ as in Theorem \ref{thm1} and any initial velocity $v_0$
(not necessarily small), if $r_0$ is chosen sufficiently large,
then the particle will turn before it reaches $x = - r_1 $.
Thus we can assure that the electron never gets closer 
than $r_1$ to  field's source 
{\em either} by taking the initial velocity sufficiently small
or by taking the cutoff $r_0$ sufficiently large. 
\section{Proof of Theorem \ref{thm1}} 
Since the electron's motion is restricted to one space dimension,
for notational simplicity we work in two-dimensional
Minkowski space with orthonormal coordinates $(t,x)$.  Our metric
gives positive norm to timelike vectors, 
and negative norm to spacelike vectors,
and the velocity of light is normalized to 1. 
The Lorentz-Dirac equation for 
a particle with mass $m$, charge $q$, and four-velocity 
$u$ is:
\beq
\lbl{ldeq}
m \frac{du^i}{d\tau} = q {F^i}_\alpha u^\alpha + \frac{2}{3} q^2
\left[\frac{d^2 u^i}{d\tau^2} + \frac{du^\alpha}{d\tau}
\frac{du_\alpha}{d\tau} u^i 
\right]
\q,
\eeq
with $u(\tau)$ the four-velocity of a charged particle
(our electron) at proper time $\tau$, and $F$ the external field 
(which depends on the Minkowski coordinates $(t,x)$ of the particle.

In two-dimensional Minkowski space, for a particle with
worldline $\tau \mapsto (t(\tau), x(\tau) )$, 
with $\tau$ proper time,
\beq
\lbl{fourvel}
u = (\gamma, v\gamma) = (\cosh \theta, \sinh \theta)
\,
\eeq
where $v = dx/dt$ is the particle's coordinate velocity,
$\gamma = \gamma(v) := (1 - v^2) ^ {-1/2}$,
and \re{fourvel} defines the ``rapidity'' parameter 
$\theta = \tanh^{-1} v$. 

Define a unit spacelike vector $w$ orthogonal to $u$:
\beq
\lbl{wdef}
w := (v\gamma, \gamma) = (\sinh \theta, \cosh \theta)
\q.
\eeq 
Physically, $w$ represents a vector pointing ``to the right''. 
At an instant at which the particle has zero velocity, $w = (0,1)$,
a rightward unit vector in the direction of the $x$-axis.

Since $du/d\tau$ is orthogonal to $u$ (as one sees from 
differentiating $u^j (\tau) u_j (\tau) = 1)$, 
it must be a multiple of $w$.
We call this multiple $A$ the {\em proper acceleration}:
\beq
\lbl{Adef}
\frac{du}{d\tau} = A w 
\q.
\eeq
From the definitions of $u$ and $w$ in terms of the rapidity $\theta$,
we see that
\beq
\lbl{Arap}
A = \frac{d\theta}{d\tau}  
\q.
\eeq

Substituting these definitions in the LD equation \re{ldeq}
and collecting terms proportional to $w$ (the sum of terms proportional 
to $u$ vanishes)
yields the following simple scalar equation, which is equivalent
to the LD equation in two spacetime dimensions:
\beq
\lbl{LDeq2}
mA = qE + \frac{2}{3} q^2 \frac{dA}{d\tau}
\q,
\eeq 
where $E$ is the scalar electric field defined by:
\beq
\lbl{scalarfield}
{F^i}_j u^j =  E w^i
\q.
\eeq
This equation defines $E$ because the antisymmetricity of $F$
implies that 
${F^i}_j u^j $  is orthogonal to $u$, and hence must be a multiple of $w$.  

In terms of the Minkowski coordinates, for the cutoff Coulomb field 
\re{largecutoff}, 
\beq
\lbl{scalarcoulomb}
E(x) = -2Q^2/3x^2 \q \mbox{for $ - r_0 < x < 0$}
\q.
\eeq
(Since the electron is coming in on the negative $x$-axis and
will be shown to turn before it reaches the origin,
we shall only be concerned with $x < 0$.)  The minus sign is because
the field is radially outward,
which is in the negative $x$-direction for $x < 0$.
That the scalar electric field for the cutoff Coulomb field 
\re{largecutoff} is of the form \re{scalarcoulomb} 
is not quite obvious, but
can be obtained from direct calculation from 
$\re{scalarfield}$.
(Basically, one is proving that in two-dimensional Minkowski space,
the electric field with respect to one orthonormal basis is
the same as that of any other orthonormal basis with the same
orientation.) 
The Lorentz force $q {F^i}_j u^j$ from such a field will attract 
an electron
toward the source proton.

We can get rid of the constants in this equations by appropriate 
normalizations.  First of all, given any charged particle,
we can always choose units of mass, charge, and time 
such that this particle has any desired mass and any desired charge 
(cf \cite{parrottbook}, Exercise 3.14.) 
We will use units in which the electron has charge $-1$
and mass $2/3$.  This changes \re{LDeq2} to:
\begin{eqnarray}
\lbl{LDeq2c}
A &=&  \frac{3}{2} E + \frac{dA}{d\tau}  
\end{eqnarray} 
To get rid of the annoying factor $3/2$,
define 
\beq
\lbl{Ebar}
\Ebar := \frac{3}{2} E
\q,
\eeq
which changes \re{LDeq2c} to 
\beq
\lbl{LDeq2d}
A = - \Ebar + \frac{dA}{d\tau}  
\q.
\eeq
Multiplying both sides by the integrating factor $e^{-\tau}$ and
rearranging yields 
$$
\frac{d (e^{-\tau} A)}{d\tau} = e^{-\tau} \Ebar
\q.
$$
Taking proper time (and also coordinate time) to be zero when
the electron enters the field at $x = -r_0$
and integrating both sides
gives:
\begin{eqnarray}
\lbl{ldsolb}
A(\tau) &=& 
A(0) e^\tau + e^\tau \int_0^\tau e^{-\tauprime} 
\Ebar(x(\tauprime)) \, d\tauprime
\\
\lbl{ldsol} 
&=&  e^\tau \int_0^\tau e^{-\tauprime} \Ebar (x(\tauprime)) \, d \tauprime
\end{eqnarray}
where  $x = x(\tauprime)$ denotes the electron's 
position at proper time $\tauprime$,
and 
the last line uses the ``no preacceleration'' assumption $A(0) = 0$. 
This is sometimes called a ``solution'' of the LD equation,
though it's really another equation for the electron's worldline because 
the right side involves the unknown worldline. 

Note that the field $\bE$ in Theorem 1 is radially outward,
in the same direction as a Coulomb field.  
Thus for $x < 0$,  the scalar field $E(x)$ is nonpositive 
and also $\Ebar$ is nonpositive. 
Thus for $ \tau > 0$, the right side of \re{ldsol}
is {\em strictly negative}: 
\beq
\lbl{accneg}
A(\tau) < 0 \q\q\q \mbox{for $\tau > 0$.} 
\eeq

When one sees this for the first time,
a typical reaction is  that a sign must have been incorrectly altered 
somewhere.
This is because the electron's acceleration is {\em in the ``wrong'' 
direction}, describing a repulsion from the proton instead of the
expected attraction.
Nevertheless, the signs in \re{ldsol} are correct.
This unexpected sign is the basic mechanism behind  Eliezer's theorem.  

Recall that the electron enters the field at position $x = -r_0$
at time zero, traveling to the right at initial velocity 
$v_0 > 0$. 
Denote its position at coordinate time $t$ by $x(t)$, 
its velocity by $v := dx/dt$, and its ``coordinate acceleration'' $A_c$
by 
\beq
\lbl{coordacc}
A_c := \frac{dv}{dt}
\q.
\eeq 
Here I am indulging in a common abuse of notation 
by writing, e.g., $x(\tau)$ for the electron's position  
at proper time $\tau$ and also $x(t)$ for the position at coordinate time $t$.
Such abuse of notation causes me to cringe as a mathematician,
but it is so common in physics that to introduce different symbols
for position considered as a functions of proper time and 
of coordinate time might 
seem pedantic and distracting. 
So long as $\tau$ is used consistently for proper time and $t$
for coordinate time, no confusion should arise. 

\newtheorem{lemma}{Lemma}
\begin{lemma} 
\lbl{lemma1}
The relation between the its acceleration $A$ and coordinate 
acceleration $A_c$ is:
$$
A_c := A \gamma^{-3} = A(1-v^2)^{3/2}
\q.
$$
\end{lemma}
Perhaps the easiest way to see this is to differentiate $v = \tanh \theta$,
recalling that $d\theta/d\tau = A$ and that $dt/d\tau = \gamma$. 

Now let's think about the implications of the unexpected sign
of the right side of \re{ldsol}.
It is obvious from \re{ldsol} that $A$ is always negative after the 
electron enters the field (traveling to the right).
Hence $A_c = dv/dt $ is also negative, which says that the particle
is {\em slowing} until it turns around (if it does).
After it turns, \re{ldsol} makes it obvious that 
it accelerates to the left exponentially with proper time,
heading toward $x = -\infty$. 

So, the issues remaining are whether the electron must turn around
before reaching the proton, and how close it can come before turning. 
These remaining issues only involve the electron's behavior before
it turns (or collides). 
So from now on we only consider times before the particle turns
(or collides). 

In resolving these issues, it may help to think of $v_0$ 
as small (though our proofs apply to any $v_0$).  
In that case, the electron's velocity $v$ which is decreasing,
is also small before it turns, 
so that $dt/d\tau = \gamma = (1 - v^2) ^ {-1/2} \approx 1.$ 
This means that there is no practical distinction between proper time
and coordinate time, 
nor between proper acceleration and coordinate acceleration.  
Of course, to write down proper proofs we have to maintain 
the distinctions (and we shall),
but for intuitive purposes we can think classically and 
talk about ``time'' or ``acceleration'' 
as if we were in a Newtonian world without time dilation.
Thinking this way may make the proofs easier to follow.

From equation \re{ldsol} for the acceleration,
we see that not only is the proper acceleration $A(\tau)$ 
strictly negative for $\tau > 0$, but its magnitude $|A(\tau)|$
can only grow (i.e., is monotonically nondecreasing) for $\tau > 0$. 
Thus for any number $r_2$ with $0 < r_2 < r_0$ 
the magnitude of the proper acceleration 
when the electron reaches $x = - r_2$ 
is a lower bound for the magnitude thereafter. 
If the electron first reaches $x = - r_2$ at proper time $\tau_2$, then
we have the following lower bound for magnitude of the proper acceleration
thereafter:
\beq
\lbl{acclbd}
|A(\tau)| \geq |A(\tau_2)| \q\q\q \mbox{for $\tau \geq \tau_2$}
\q.
\eeq 

Now we can describe the idea behind the remainder 
of the proof.  
We want to show that if the electron's initial velocity $v_0$ is small enough,
then it cannot reach $x = -r_1$.
Assume the contary, to obtain a contradiction.
 
We have a lower bound for the magnitude of the acceleration
on a trip from $x = -r_2$ to $x = -r_1$, 
and by integration, this will imply a lower bound for 
decrease in velocity on this trip.  
If this lower bound for the decrease in velocity is larger than
the initial velocity $v_0$, then the particle must turn before 
it reaches $x = -r_1$, so it never reaches $x = -r_1$. 
We shall show that for sufficiently small $v_0$, this lower bound
for the decrease in velocity is indeed larger than $v_0$.

Before we can carry this out, we need to sharpen the above
lower bound \re{acclbd} to remove its dependence on $\tau_2$.
The problem is that $\tau_2$ depends on the electron's motion,
and the motion depends on the initial velocity $v_0$.
Later we will need to take $v_0$ arbitrarily small,
so to carry out the above program, we'll need a lower bound for
$A_c$ which explicitly displays its dependence on $v_0$. 
The following lemma accomplishes this.
\begin{lemma}
\lbl{lemma3a} 
Given $r_0$ and $r_1$ as described in the statement of Theorem \ref{thm1}, 
choose $r_2$ with $ r_1 < r_2 < r_0$ such that 
the electron reaches $x = - r_2$ before turning.
Let $\tau_2$ denote the proper time at which this occurs. 

Let
$$ K := \int_{-r_0}^{-r_2} |E(x)| \, dx \ > \ 0
\q.
$$ 
Then the electron's proper acceleration $A$ at proper time $\tau_2$ 
satisfies
\beq
\lbl{intermed}
|A(\tau_2)| \geq \frac{1}{v_0\gamma(v_0)} K
\q,
\eeq
and its coordinate acceleration $A_c$
when it reaches $x = -r_2$ at proper time $\tau = - \tau_2$ 
satisfies:
\beq
\lbl{indepbnd}
|A_c| \geq 
 \frac{(1 - v^2_0)^2}{v_0} K
\eeq 
\end{lemma}
\noindent
{\bf Proof of Lemma \ref{lemma3a}:} 
Let $x(\tau)$ denote the electron's position at proper time $\tau$,
let $t$ represent coordinate time,
and $v$ the electron's velocity.
For future use, note that 
$$
\frac{dx}{d\tau} = \frac{dx}{dt} \frac{dt}{d\tau}
= v \gamma (v)
\q,
$$ 
and that $v \mapsto v \gamma(v)$ is monotonically increasing.  
At any point on the electron's trip from $x = -r_0$ to 
$x  = - r_2$,
we have that $v \leq v_0$, so
$$
\frac{dx}{d\tau} \leq v_0\gamma(v_0)
\,
$$
equivalently,
$$
\frac{1}{dx/d\tau} \geq \frac{1}{v_0\gamma(v_0)}
\q.
$$ 

The ``solution'' \re{ldsol}  of the LD equation 
tells us that
\begin{eqnarray}
|A(\tau_2)| &=& 
e^{\tau_2} \int_0^{\tau_2}  -\Ebar(x(\tauprime)) e^{-\tauprime}
 \, d\tauprime 
\nonumber\\
&\geq & \int_0^{\tau_2}  -\Ebar(x(\tauprime)) \, d\tauprime 
\nonumber\\
&=&  \int_0^{\tau_2} -\Ebar(x(\tauprime)) 
\frac{dx/d\tauprime}{dx/d\tauprime} \, d\tauprime 
\nonumber\\
&\geq&  
\frac{1}{v_0 \gamma(v_0)} \int_{-r_0}^{-r_2} -\Ebar(x) 
\, dx 
\nonumber\\ 
&=& 
\frac{1}{v_0 \gamma(v_0)} K 
\nonumber
\end{eqnarray}
This proves inequality  \re{intermed} in the statement of the Lemma.

The final inequality \re{indepbnd} follows immediately from 
\re{intermed} combined with the relation $A_c = A \gamma^{-3}$ 
(Lemma \ref{lemma1}).  Denoting by $v_2$ the electron's
velocity at proper time $\tau_2$, we have at that time:
$$
A_c = A(\tau_2) \gamma(v_2)^{-3} \geq  
\frac{1}{v_0 \gamma(v_0)} K \gamma(v_0)^{-3}  = \frac{(1 - v^2_0)^2}{v_0} K 
\q.
$$
This completes the proof of Lemma \ref{lemma3a}.

The rest of the proof of Theorem 1 is easy.  
Recall that we are assuming that the electron
enters the field at $x = - r_0$ at coordinate time $0$. 
We are also assuming, to obtain a contradiction, that it reaches
$x = - r_1$.  Let $t_1$ denote the coordinate time at which it first
reaches $x = - r_1$, and let $v_1$ be its velocity at that time. 
Let $t_2$ be the coordinate time at which it first reaches $x = - r_2$.
Then from Lemma \ref{lemma3a}, we have 
\begin{eqnarray*} 
v_0  - v_1 &=&   \int_{t_l}^{0}  \frac{dv}{dt} \,dt \nonumber\\ 
&=& \int_0^{t_1} - \frac{dv}{dt}\, dt \nonumber\\
&=& \int_0^{t_1} |A_c| \, dt \nonumber\\ 
&\geq& 
\int_{t_2}^{t_1} |A_c| \, dt \nonumber\\ 
&\geq&  (t_1- t_2) \frac{(1 - v^2_0)^2}{v_0} K 
\end{eqnarray*}
No particle can exceed the velocity of light, so the time $t_1 - t_2$
to go from $x = -r_2$  to $x = -r_1$ 
must be at least as large as the corresponding
distance $r_2 - r_1$.  Hence
\beq
\lbl{velbound}
v_0 \geq 
(r_2 - r_1) \frac{(1 - v^2_0)^2}{v_0} K 
\q,
\eeq 
which is equivalent to
\beq
\lbl{contra}
\frac{v^2_0}{ (1 - v^2_0)} \geq K(r_2 - r_1)
\q.
\eeq
Taking $v_0$ sufficiently small violates this last inequality.
This contradiction shows that the premise of the inequality, i.e., that
the electron does reach $x = - r_1$, must be false.
This completes the proof of Theorem \ref{thm1}.
\section{Another variant of Eliezer's theorem} 
Theorem \ref{thm1} is surprisingly general.
It applies to virtually any radially outward electric field
which eventually vanishes.
However, it does require that the initial velocities be small.

If we are willing to return to a cutoff Coulomb field \re{largecutoff},
we can prove the following version of Eliezer's theorem which applies to 
any arbitrary initial velocity if the cutoff is chosen sufficiently large.  
\begin{theorem}
\lbl{thm2}
Consider a  Coulomb field \re{largecutoff} cutoff at $r = r_0$, 
and consider an electron entering the field from the left at $x = -r_0$ 
with velocity $v_0 > 0$ and zero acceleration at the time of entry. 

Let a distance $r_1$ be given.
Then given any $v_0$, there exists a cutoff $r_0$ (which can be
taken arbitrarily large) such that the electron never gets closer
than $r_1$ to the ``proton'' source at the origin. 
Instead, it turns before 
it reaches $x = -r_1$ and thereafter travels back toward $x = -\infty$
with proper acceleration exponentially increasing with proper time.
\end{theorem} 
The ideas of the proof of this are the same as the ideas discussed
in Theorem \ref{thm1}.
The proof of Theorem \ref{thm2} follows the proof of Theorem \ref{thm1}
through Lemma \ref{lemma3a}. 
However, in the present case the constant $K$ in Lemma \ref{lemma3a}
can be explicitly computed as:
$$
K := \int_{-r_0}^{-r_2} |E(x)|\, dx = \left[ \frac{1}{r_2} - 
\frac{1}{r_0} \right]
\q.
$$
Lemma \ref{lemma3a} holds for {\em any} $r_2$ with $r_2 < r_0$ 
and such that the electron reaches $ x = - r_2$.
Hence what was actually proved may be rewritten as 
\begin{lemma}
\lbl{lemma3b}
Let $x(t)$ denote the electron's position at coordinate time $t$.
Then its coordinate acceleration  $A_c(t)$ at that time
satisfies: 
\beq
\lbl{coulbound}
|A_c (t)| \geq 
 \frac{(1 - v^2_0)^2}{v_0}
\left[-\frac{1}{x(t)} - \frac{1}{r_0} \right] 
\q.
\eeq 
\end{lemma}
(The sign of $- 1/x(t)$ may look suspicious,
but recall that $x(t)$ is negative.  This term corresponds
to the positive term $1/r_2$ in the preceding equation.) 

Suppose (to obtain a contradiction for sufficiently large $r_0$) 
that the electron reaches $x = - r_1$ before turning, and let 
let $v_1$ denote its velocity at that time.
Integrating \re{coulbound} gives a lower bound for 
the decrease in velocity between $ x = - r_0$ and $ x = - r_1$:
\begin{eqnarray*} 
\lbl{coulvelbnd}
v_0 - v_1 
&=& \int_{t_1}^0 A_c(t) \, dt \\
&=& \int_0^{t_1} |A_c(t)| \, dt \\
&\geq&  
 \frac{(1 - v^2_0)^2}{v_0} \int_0^{t_1} 
\left[- \frac{1}{x(t)} - \frac{1}{r_0} \right] \, dt \\
&=&
 \frac{(1 - v^2_0)^2}{v_0} \int_0^{t_1} \left[
- \frac{1}{x(t)} - \frac{1}{r_0}\right] \frac{dx/dt}{dx/dt} \, dt \\
&\geq&
 \frac{(1 - v^2_0)^2}{v_0} \frac{1}{v_0}
 \int_{-r_0}^{-r_1}
\left[ - \frac{1}{x} - \frac{1}{r_0}  \right] \, dx \\
&=& 
 \frac{(1 - v^2_0)^2}{v^2_0} 
\left[ \log(r_0/r_1) - \frac{r_0 - r_1}{r_0} \right] 
\q.
\end{eqnarray*}
Hence the particle will turn before it reaches $x = - r_1$
if $r_0$ is large enough so  that the right side is at least $v_0$.
This completes the proof of Theorem \ref{thm2} 

Notice that the right side of the last inequality can also be made
arbitrarily large by taking $r_1$ sufficiently small.
This proves the following variant of Eliezer's Theorem
which assures that under the hypotheses of Theorem 2, for
fixed  $r_0$ and $v_0$, {\em all} solutions are runaway
(in the ``wrong'' direction).

\begin{theorem}
\lbl{thm3} 

Let a  Coulomb field \re{largecutoff} cutoff at $r = r_0$ be given, 
and consider an electron entering the field from the left at $x = -r_0$ 
with positive velocity and zero acceleration at the time of entry. 
Then the electron cannot approach arbitrarily close 
to the field's  ``proton'' source at the origin.

Instead, the electron  turns before 
it reaches  the origin and thereafter travels back toward $x = -\infty$
with proper acceleration exponentially increasing with proper time.  
\end{theorem}
This result was proved in \cite{parrottfp}, using similar techniques. 
The basic ideas of the proof, which go back to Eliezer \cite{eliezer},
have been refined over the years by various authors.  The  treatment 
above was strongly influenced by \cite{hsing} (which unfortunately
was never published) and \cite{hsing2}. 
%

\end{document}